\documentclass[twocolumn, preprint,aps,prb,superscriptaddress,showpacs,10pt,floatfix,twocolo,amsmath]{revtex4-2}
\usepackage{epsfig}
\usepackage{amsmath}
\usepackage{bm}
\usepackage{here}
\usepackage{float}
\usepackage{graphicx}
\usepackage[linkcolor=black,urlcolor=blue,citecolor=blue,colorlinks=true]{hyperref}
\usepackage{xcolor}

\usepackage[normalem]{ulem}

\begin{document}
	
\title{Complexity in the hybridization physics revealed by depth-resolved photoemission spectroscopy of single crystalline novel Kondo lattice systems, CeCuX$_2$ (X = As/Sb)}

\author{Sawani Datta,$^1$ Ram Prakash Pandeya,$^1$ Arka Bikash Dey,$^2$ A. Gloskovskii,$^2$ C. Schlueter,$^2$ T. R. F. Peixoto,$^2$ Ankita Singh,$^1$ A. Thamizhavel,$^1$ and Kalobaran Maiti}
\altaffiliation{Corresponding author: kbmaiti@tifr.res.in}
\affiliation{Department of Condensed Matter Physics and Materials Science, Tata Institute of Fundamental Research, Homi Bhabha Road, Colaba, Mumbai-400005, India.\\
$^2$Deutsches Elektronen-Synchrotron DESY, 22607 Hamburg, Germany.}
	
\begin{abstract}
We investigate the electronic structure of a novel Kondo lattice system CeCuX$_2$ (X = As/Sb) employing high resolution depth-resolved photoemission spectroscopy of high quality single crystalline materials. CeCuSb$_2$ and CeCuAs$_2$ represent different regimes of the Doniach phase diagram exhibiting Kondo-like transport properties and CeCuSb$_2$ is antiferromagnetic ($T_N \sim$ 6.9 K) while CeCuAs$_2$ does not show long-range magnetic order down to the lowest temperature studied. In this study, samples were cleaved in ultrahigh vacuum before the photoemission measurements and the spectra at different surface sensitivity establish the pnictogen layer having squarenet structure as the terminated surface which is weakly bound to the other layers. Cu 2$p$ and As 2$p$ spectra show spin-orbit split sharp peaks along with features due to plasmon excitations. Ce 3$d$ spectra exhibit multiple features due to the hybridization of the Ce 4$f$/5$d$ states with the valence states. While overall lineshape of the bulk spectral functions look similar in both the cases, the surface spectra are very different; the surface-bulk difference is significantly weaker in CeCuAs$_2$ compared to that observed in CeCuSb$_2$. A distinct low binding energy peak is observed in the Ce 3$d$ spectra akin to the scenario observed in cuprates and manganites due to the Zhang-Rice singlets and/or high degree of itineracy of the conduction holes. The valence band spectra of CeCuSb$_2$ manifest highly metallic phase. In CeCuAs$_2$, intensity at the Fermi level is significantly small suggesting a pseudogap-type behavior. These results bring out an interesting scenario emphasizing the importance and subtlety of hybridization physics underlying the exoticity of this novel Kondo system.
\end{abstract}
	
\maketitle
	
\section{Introduction}
	
Rare earth-based compounds exhibit several exotic properties due to the local character of the rare earth 4$f$ states and their hybridization with the conduction electrons \cite{review,Parlebas,book}. In Ce-based materials, the occupied part of the 4$f$ band appears close to the Fermi level, $\epsilon_F$ leading to strong hybridization of these states with the conduction electronic states and thereby these 4$f$ electrons gain significant itineracy in addition to their local character. Such dual properties of the Ce 4$f$ electrons leads to plethora of exotic properties and are among the most studied systems in this class \cite{review, Parlebas, book, patil-213bulk, Sengupta1, CeCuX2-1,CeCuX2-2, thamizh, patil_CEF}. Moreover, Ce 4$f$ orbitals have larger radial extension than the heavier rare-earth 4$f$ states and hence, Ce 4$f$ electrons are relatively more exposed to the crystal field that gives rise to additional complexity in the physics of these materials \cite{Weschke, Allen, patil_CEF}. It has been found that the properties of the conduction electrons (correlation induced effect, spin-orbit coupling, etc.) also play key role in the exoticity of these materials \cite{patil-SOC}. For example, Ce$_2$CoSi$_3$ is a kondo material and does not show long-range magnetic order till the lowest temperature studied\cite{Gordon}, while Ce$_2$RhSi$_3$ belonging to the same class show antiferromagnetic (AFM) ground state\cite{Chevalier, Szlawska} and considered to be an example of quantum critical behavior with spin density wave ground state \cite{patil-213bulk}.

\begin{figure}
\centering
\includegraphics[width=0.5\textwidth]{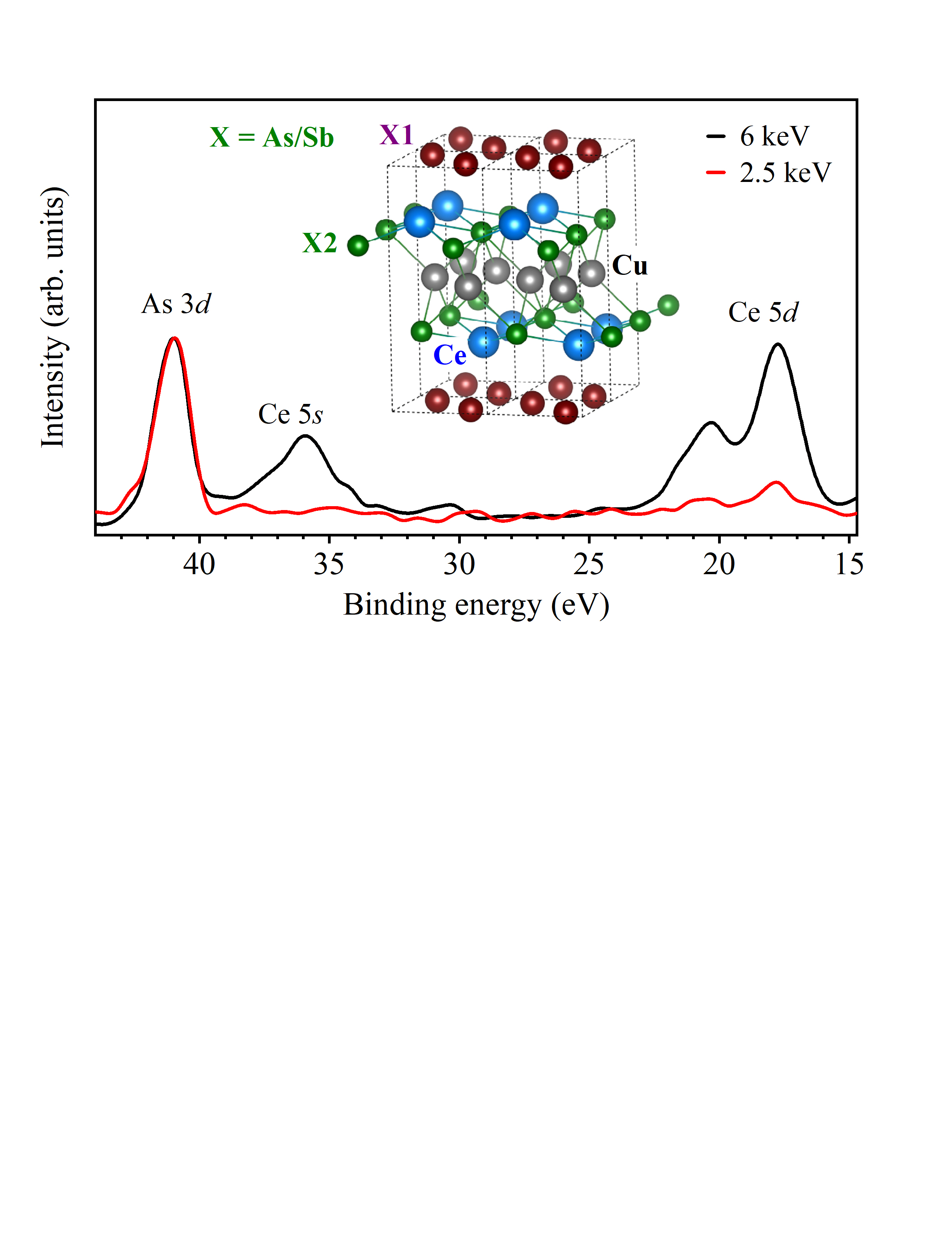}
\vspace{-44ex}
\caption{As 3$d$, Ce 5$s$ and Ce 5$d$ survey spectra of CeCuAs$_2$ collected using 2.5 keV (solid circles) and 6 keV (open circles) photon energies. Inset shows the crystal structure CeCuX$_2$; X = pnictogen, As/Sb.}
\label{Fig1-cleave}
\end{figure}

Here, we study the electronic structure of single crystalline CeCuX$_2$ (X = As/Sb) employing high resolution depth-resolved photo-electron spectroscopy (DRPES) using hard $x$-rays, where the substitution is done at the pnictogen sites. These materials form in ZrCuSi$_2$-type layered tetragonal structure (space group $P4/nmm$) as shown in the inset of Fig. \ref{Fig1-cleave} \cite{ Sengupta1,thamizh}. The pnictogen atoms X (X = As, Sb) have two different Wyckoff positions; X1 atoms form a squarenet structure and weakly coupled to Ce layers. X2 is strongly hybridized with Cu atoms along with reasonably strong Ce-X2 bonds. Larger atomic radius of Sb atoms leads to a slightly larger lattice parameters; $a$ = 4.018 and 4.337 \AA, and $c$ = 10.104 and 10.233 \AA\ for X = As and Sb, respectively. The Ce-Cu, Ce-X1 and Ce-X2 bondlengths in CeCuSb$_2$ (= 3.3615, 3.3459 and 3.2297 \AA) are larger than those in CeCuAs$_2$ (= 3.2992, 3.1569 and 2.9279 \AA). However, Cu-As2 (= 2.7718 \AA) bond is weaker than Cu-Sb2 (= 2.6686 \AA) bond. Since, X2-Cu-X2 layers are the conduction layers in these materials, such change in Cu-X2 bondlength has significant implication in the properties of the conduction electrons. Shorter Cu-Sb2 bondlength leads to a higher degree of metallicity of CeCuSb$_2$ as also observed in the bulk properties study of these samples. The conduction electrons in As2-Cu-As2 layers are expected to be relatively less itinerant and are more strongly coupled to the Ce 4$f$ states than those in CeCuSb$_2$.


The bulk properties of this class of materials have been studied extensively and most of the candidates show long range magnetic order \cite{CeCuX2-1,CeCuX2-2}. Among the two systems studied here, CeCuSb$_2$ exhibits AFM ground state with Ne\'{e}l temperature, 6.9 K \cite{thamizh}, whereas CeCuAs$_2$ does not show long-range magnetic order down to the lowest temperature studied \cite{Sengupta1, Sengupta2}. Resistivity ($\rho$) of both the materials increases logarithmically with cooling from room temperature similar to a Kondo system. A maxima is observed around 23 K (for J $||$ [100]) in CeCuSb$_2$ and subsequently, $\rho$ gradually decreases at lower temperatures \cite{thamizh}. In CeCuAs$_2$, the resistivity maxima is observed at 20 K and the slope of the curve decreases at lower temperatures; such unusual behavior at low temperatures is not well understood yet \cite{Sengupta1}. The in-plane resistivity of CeCuAs$_2$ is much higher than that for CeCuSb$_2$, which suggests that CeCuSb$_2$ is a better electronic conductor. The AFM ground state along with Kondo behavior in CeCuSb$_2$ places this material in the spin density wave (SDW) type quantum critical regime such as Ce$_2$RhSi$_3$, where Ruderman-Kittel-Kasuya-Yosida (RKKY) interaction mediates the magnetic order. In CeCuAs$_2$, the Kondo behavior dominates giving rise to a paramagnetic ground state as observed in Ce$_2$CoSi$_3$. The uniqueness of these materials is that the tuning of the properties of conduction electrons occurs via substitution at the pnictogen site in contrast to the earlier cases where the transition metal site was used for substitution. Moreover, the pnictogen layer forming the squarenet structure hosts Dirac fermions and is in proximity to the Ce-layer while the pnictogen layer on the other side is strongly bonded to Ce and Cu layers. Thus, these materials provide a novel playground to study the complex physics involving behavior of Ce 4$f$ electron in proximity to topological states with varying Kondo coupling strength. We observe interesting features in the DRPES results revealing puzzling scenario in this system.

\section{Experiment}
	
High-quality single crystals of CeCuSb$_2$ and CeCuAs$_2$ were prepared by flux method using Sb and As-Sn flux, respectively. Elemental analysis of the prepared samples was done by the energy dispersive analysis of $x$-rays. Good crystallinity of the samples was confirmed by the Laue diffraction measurements. The hard $x$-ray photoemission spectroscopy (HAXPES) measurements were performed at the P22 beamline of Petra III, DESY, Hamburg, Germany. We have used a high-resolution Phoibos electron analyzer; the energy resolution at 6 keV photon energy is close to 150 meV. The resolution becomes close to 200 meV at other photon energies. All the measurements were carried at a pressure lower than $\sim 10^{-10}$ Torr, immediately after cleaving the samples in ultrahigh vacuum employing a top-post removal method. The sample temperature was varied using an open cycle helium cryostat. To tune the surface sensitivity of the measurement technique, we have used three well-separated incident photon energies 2.5 keV, 6 keV and 8 keV.


In order to capture the properties of various features in the valence band, we calculated the electronic band structure were using full potential linearized augmented plane wave method (FLAPW) as implemented in the Wien2k software \cite{Blaha}. The Perdew-Burke-Ernzerhof generalized gradient approximation (GGA) \cite{Perdew} was used for the calculation of the density functionals \cite{Hohen}. We considered electron correlation strength, $U$ = 7.5 eV for 4$f$ electrons and spin-orbit coupling for the calculations. Lattice parameters were fixed at the experimentally derived values for our calculations \cite{thamizh, Sengupta1}.

\section{Results and discussion}

In Fig. \ref{Fig1-cleave}, we show the survey scans consisting of As 3$d$, Ce 5$s$ and Ce 5$d$ core level spectra of CeCuAs$_2$ collected using 2.5 keV and 6 keV photon energies. The escape depths of these photoelectrons at 6 keV and 2.5 keV photon energies are close to 40 \AA\ and 26 \AA, respectively \cite{escapedepth}. Hence, the surface sensitivity of the experiment will be significantly larger at 2.5 keV photon energy compared to the 6 keV case. In the figure, we normalized the spectral intensities by the intensity of the As 3$d$ peak height. This shows that the intensities of Ce core level peaks are significantly larger in the bulk sensitive 6 keV data. The intensity of these core level peaks depends essentially on (i) the photoemission cross-section and (ii) the number of electrons at the corresponding energy level. The cross-section ratio of As 3$d$ and Ce 5$d$ at 2.5 keV and 6 keV are very similar; $I(As 3d)/I(Ce 5p)$ = 2.78 (2.5 keV) and 2.66 (6 keV) using atomic photoemission cross-sections \cite{Yeh}. The occupancy of the energy levels in the ground state is the same in both the cases. Thus, the huge change in intensity observed in the experimental data in Fig. \ref{Fig1-cleave} can be attributed to the surface sensitivity of the technique and As layer appears to be the terminated surface of the sample.

From the crystal structure shown in the inset of Fig. \ref{Fig1-cleave}, it is clear that As1 layers forming the squarenet structure are relatively far away from the Ce-As2-Cu-As2-Ce quintuple layers and are weakly bonded to the Ce-layers on both sides of the As1-plane. The covalent bonding of As2-layer with Ce and Cu layers is strong. From these structural parameters and the inter-layer bonding, the cleaving of the sample is expected to occur at the As1-layer leading to As1-layer or Ce-layer as the terminated surface. The experimental data shown in the figure suggests that the terminated layer in CeCuAs$_2$ is the As1 layer for the cleaved sample used for our experiments. A similar scenario is observed in CeCuSb$_2$ \cite{sawani}.


\begin{figure}
\centering
\includegraphics[width=0.5\textwidth]{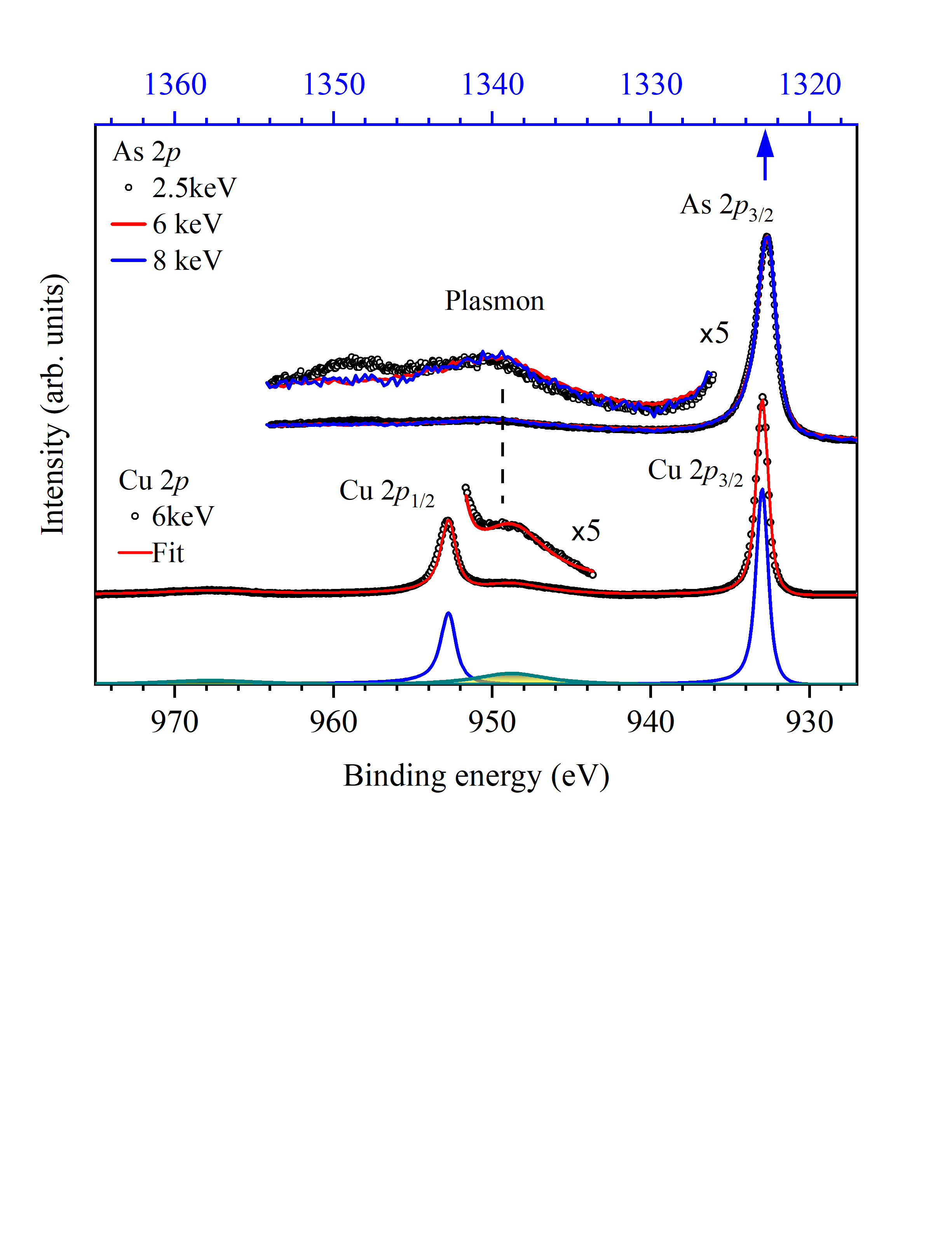}
\vspace{-32ex}
\caption{As 2$p$ ($x$-scale at the top axis) and Cu 2$p$ ($x$-scale at the bottom axis). Signature of plasmon feature is shown in an enlarged intensity scale. As 2$p_{3/2}$ peak at 2.5 keV (open circles), 6 keV (red line) and 8 keV (blue line) photon energy overlap on each other well. Line superimposed over the Cu 2$p$ spectrum is the fit curve; the component peaks are shown in the bottom panel.}
\label{Fig2-Cu2pAs2p}
\end{figure}

As 2$p_{3/2}$ spectra collected at 2.5 keV, 6 keV and 8 keV are shown in Fig. \ref{Fig2-Cu2pAs2p} exhibiting a sharp feature at 1323 eV binding energy in all the cases along with an asymmetry towards higher binding energies. This type of asymmetry in the spectral lineshape may arise due to the low energy excitations across the Fermi level along with the core level photoemission. The spectral functions around the main peak appear similar at all the photon energies probed. This suggests that the influence of surface-bulk difference, if there is any, on the deep core level, As 2$p$ photoemission is within the experimental limits. This is not unusual as these deep core levels are less exposed to the crystal field and the behavior is essentially like atomic levels. The scenario is similar for 2$p_{1/2}$ region and not shown here for clarity of presentation. In addition to the main peak, we observe two broad humps at about 1339 eV and 1349 eV binding energies; the features are shown in an enlarged intensity scale in the same figure for clarity. These features are attributed to the plasmon excitations along with the photo-excitations of the As 2$p$ core electrons. The intensity of the 1349 eV peak is larger in the 2.5 keV data compared to the intensity of the other one suggesting it's link to the surface electronic structure.

In the middle panel of Fig.\ref{Fig2-Cu2pAs2p}, we show the Cu 2$p$ core-level spectrum collected using 6 keV photon energy. Both the spin-orbit split features of Cu 2$p$ photoemission signal exhibit an asymmetric single peak structure as observed in the As 2$p$ case. The experimental data can be fitted using an asymmetric Gausian-Lorentzian function. The simulated envelope (red line) is superimposed over the experimental curve; the individual components are shown in the lower panel of Fig.~\ref{Fig2-Cu2pAs2p}. The sharp intense peaks of the spin-orbit split Cu 2$p$ features and the absence of satellites suggests close to monovalency of Cu with 3$d^{10}$ electronic configuration \cite{chainani}. In addition, there is a broad hump at about 16 eV away from the main spin-orbit split peaks as also observed in the As 2$p$ spectra which substantiates the attribution of this peak to plasmon excitations \cite{Biswas}.

\begin{figure}
\centering
\includegraphics[width=0.5\textwidth]{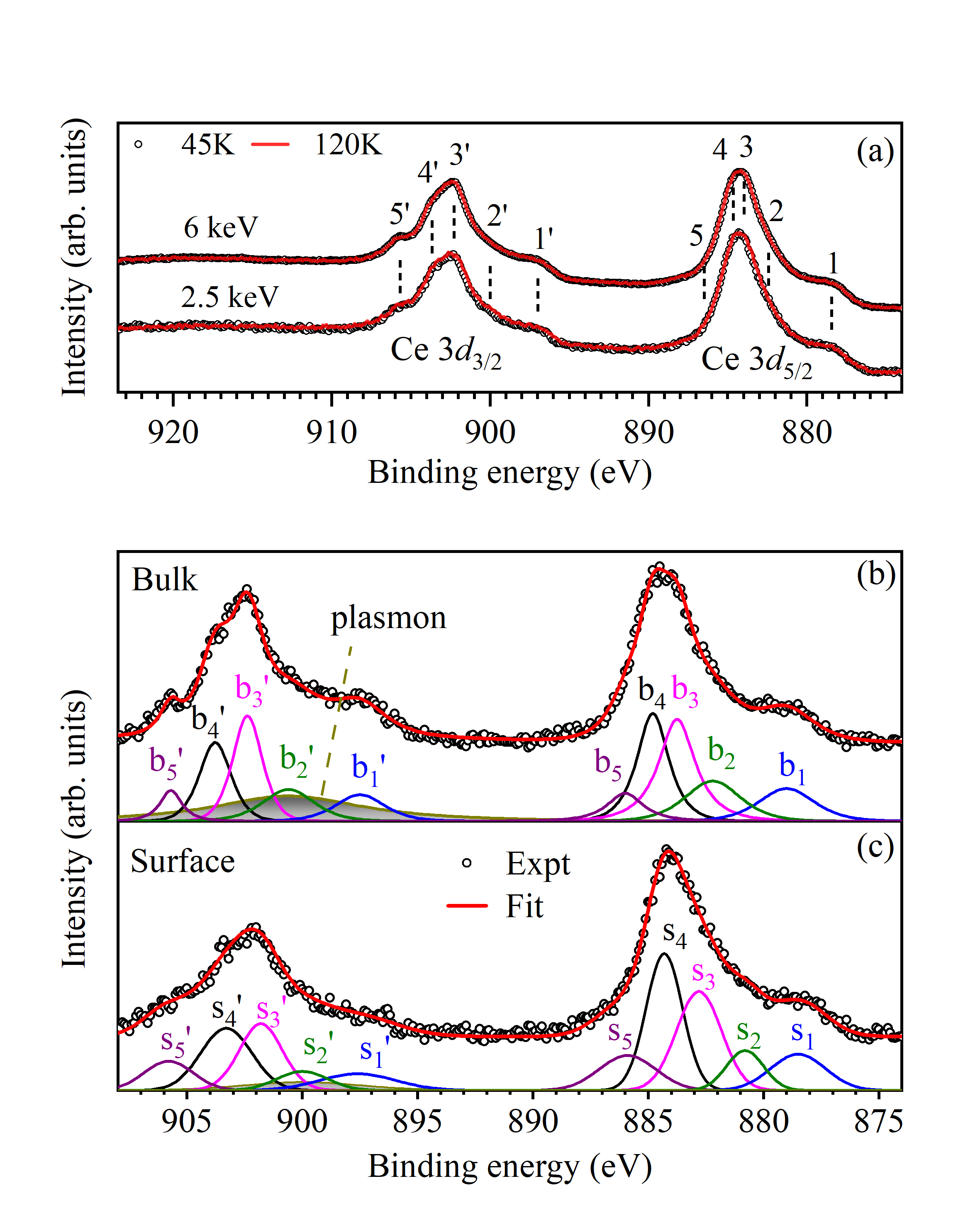}
\caption{(a) Ce 3$d$ spectra of CeCuAs$_2$ collected at 45 K (open circles) and 120 K (solid line) using 6 keV (upper panel) and 2.5 keV (lower pannel) photon energies. The spin-orbit split 3$d_{5/2}$ and 3$d_{3/2}$ features are marked by the 1,2, ... etc. and 1', 2', ... etc. respectively. The extracted (b) bulk and (c) surface spectra (open circle). Lines superimposed over the experimental data are the fit curves. The component peaks are shown in the bottom panel. The plasmon feature is shown only in the bulk spectrum for clarity; the surface spectrum also has similar feature with relatively weaker intensity.}
\label{Fig3-Ce3d}
\end{figure}
	
In Fig. \ref{Fig3-Ce3d} we study the Ce 3$d$ spectra obtained at varied experimental conditions. The experimental data are taken using 6 keV and 2.5 keV photon energies; the data at 45 K and 120 K sample temperatures are superimposed over each other in Fig. \ref{Fig3-Ce3d}(a) exhibiting identical lineshape. There are several features in the spectral regions corresponding to spin-orbit split 3$d_{5/2}$ and 3$d_{3/2}$ photoemissions. The signature of distinct peaks are marked by 1, 2, ... for 3$d_{5/2}$ signal and those in 3$d_{3/2}$ signal are marked by the primed numbers, 1$^\prime$, 2$^\prime$, .... Relative intensity of the features at different photon energies are somewhat different. For example, the intensity of the peak, 4$^\prime$ in the 6 keV data is enhanced in the 2.5 keV data relative to the intensity of the peak, 3$^\prime$, while the intensity of 5$^\prime$ is reduced. Relative intensities of the features, 1 and 2 and their primed counterparts also show subtle changes in intensity with the change in photon energy. Since the surface sensitivity of the technique enhances at 2.5 keV photon energy relative to that at 6 keV, the spectral changes observed at the two photon energies suggest different surface and bulk electronic structures.

The spectral functions corresponding to the surface and bulk electronic structures are extracted using the surface sensitivity of the technique following the well-established procedures\cite{surface-bulk}. Considering that the different surface electronic structure appears from the top few layers of thickness, $d$ and electron escape depth, $\lambda$, the spectral intensity, $I(\epsilon)$ can be expressed as, $I(\epsilon) = \int_{0}^{d}I^s(\epsilon)e^{-x/\lambda}dx + \int_{d}^{\infty}I^b(\epsilon)e^{-x/\lambda}dx$, where $I^s(\epsilon)$ and $I^b(\epsilon)$ are the surface and bulk spectra. From the above equation the spectral intensity can be derived as, $I(\epsilon) = I^s(\epsilon)(1 - e^{-d/\lambda}) + I^b(\epsilon)e^{-d/\lambda}$. Since the core levels are fully occupied, the area under the spectra representing the spectral intensity will be same at all photon energies. Thus, we normalized the experimental spectra at 2.5 keV and 6 keV by the integrated area under the curve. Considering that the universal curve is applicable to scale the escape depth\cite{escapedepth} and the spectral intensity is positive at all energies, we extracted the surface and bulk spectra functions using the above equation, and the spectra at 2.5 keV and 6 keV data.

The extracted bulk and surface 3$d$ spectra are shown in Fig.\ref{Fig3-Ce3d}(b) and (c), respectively; we have not shown the higher binding energy regime for better clarity as there are no distinct features in that region of the experimental data. To find out the peaks associated with different final states of the Ce 3$d$ photoemission, we have simulated both the spectra considering a Voigt function for each of the distinct feature in the experimental data as marked. The branching ratio (intensity ratio of the spin-orbit split features) are kept fixed to their degeneracy. The simulated data, shown by the red lines, are superimposed over the experimental extracted data in the figure and the individual peaks are shown in the lower panel. Peaks in the surface and bulk spectra are marked by s's and b's, respectively. Simulated data show a good representation of the experimental spectra.
	
In a correlated system, the eigenstates of the final state Hamiltonian will be different from those of the initial state Hamiltonian due to the presence of the photohole interacting with the other electrons. This allows transition to many final states leading to multiple features in the photoemission signal. The scenario can be represented by the configuration interaction model where each electronic configuration corresponds to a particular level of core-hole screening due to the hopping of electrons to the photoemission sites in such systems. The scenario is well captured by the Gunnarsson and Schonhammer's (GS) approach using the single impurity Anderson model (SIAM) for the calculation of Ce 3$d$ final state \cite{gun}. According to the GS calculation, three final states configurations provide major contributions in the Ce 3$d$ core level spectrum: (a) $f_2$ feature is the $well-screened$ final state, where the 3$d$ core hole is screened via hopping of a neighbor electron to the Ce 4$f$ level \cite{nickelates_kbm,RMP_Fujimori}. The presence of this feature depends on the hybridization of Ce 4$f$ level with the valence states allowing transfer of electron. (b) $f_1$ peak corresponds to the $poorly-screened$ state. Here, the final states possesses 4$f^1$ electronic configuration and the core hole is not screened. (c) $f_0$ feature relates to the final state where the Ce 4$f$ electron has hopped to the valence band leading to Ce$^{4+}$ valency; this feature is often linked to the Kondo singlet which is a quantum entangled state of the Ce 4$f$ electron with the conduction electrons \cite{gun_f0}. We did not observe signature of this feature in the present system. Each configuration consists of several multiplets; distinct signature of the multiplets depends on the eigenenergy separation of the multiplets.
	
In the Ce 3$d$ bulk spectra presented in Fig.\ref{Fig3-Ce3d}(b), the Ce 3$d_{5/2}$ peaks are denoted by b's and the spin-orbit split Ce 3$d_{3/2}$ spectrum consists of peaks b$^\prime$'s. The spin-orbit splitting is found to be around 18.8 eV. From the binding energy considerations, we ascribe the b$_2$ and b$_2^\prime$ peaks to $f_2$ final state. The features, b$_3$, b$_4$ and b$_5$ (b$_3^\prime$ b$_4^\prime$ and b$_5^\prime$ in the 3$d_{3/2}$ region) peaks to $f_1$ final state. The three-peak structure of the $f_1$ final states is observable due to the multiplet splitting. This description is consistent with the calculated results within the Anderson impurity model \cite{chainani}. We also observe a broad hump in the higher binding energy side of the main features representing the plasmon-induced loss feature.

The analysis of the surface spectrum is shown in Fig.\ref{Fig3-Ce3d}(c). Here, Ce 3$d_{5/2}$ and Ce 3$d_{3/2}$ peaks are denoted by s's and s$^\prime$'s. Plasmon feature is weaker in the surface spectrum than that in the bulk. Similar to the bulk case, s$_2$ and s$_2^\prime$ represent the well-screened $f_2$ features. There are three multiplet features in the $f_1$ signal denoted by s$_3$, s$_4$, s$_5$ for 3$d_{5/2}$ signal and s$_3^\prime$, s$_4^\prime$, s$_5^\prime$ for 3$d_{3/2}$ signal. The intensity of the $f_2$ peak relative to the $f_1$ intensity is weaker in the surface spectrum compared to that in the bulk. The binding energy of the surface peak is reduced by about 0.3 eV from the corresponding bulk peaks. These observations suggest that Ce is slightly less positive at the surface relative to the bulk Ce-valency. This is consistent with the fact that the bulk Ce states are expected to have stronger hybridization with the valence states leading to larger extended character while the surface states will be more local\cite{Laubschat}. This may also be a reason for the higher $f_2$ intensity in the bulk spectra.
	
\begin{figure}
\centering
\includegraphics[width=0.5\textwidth]{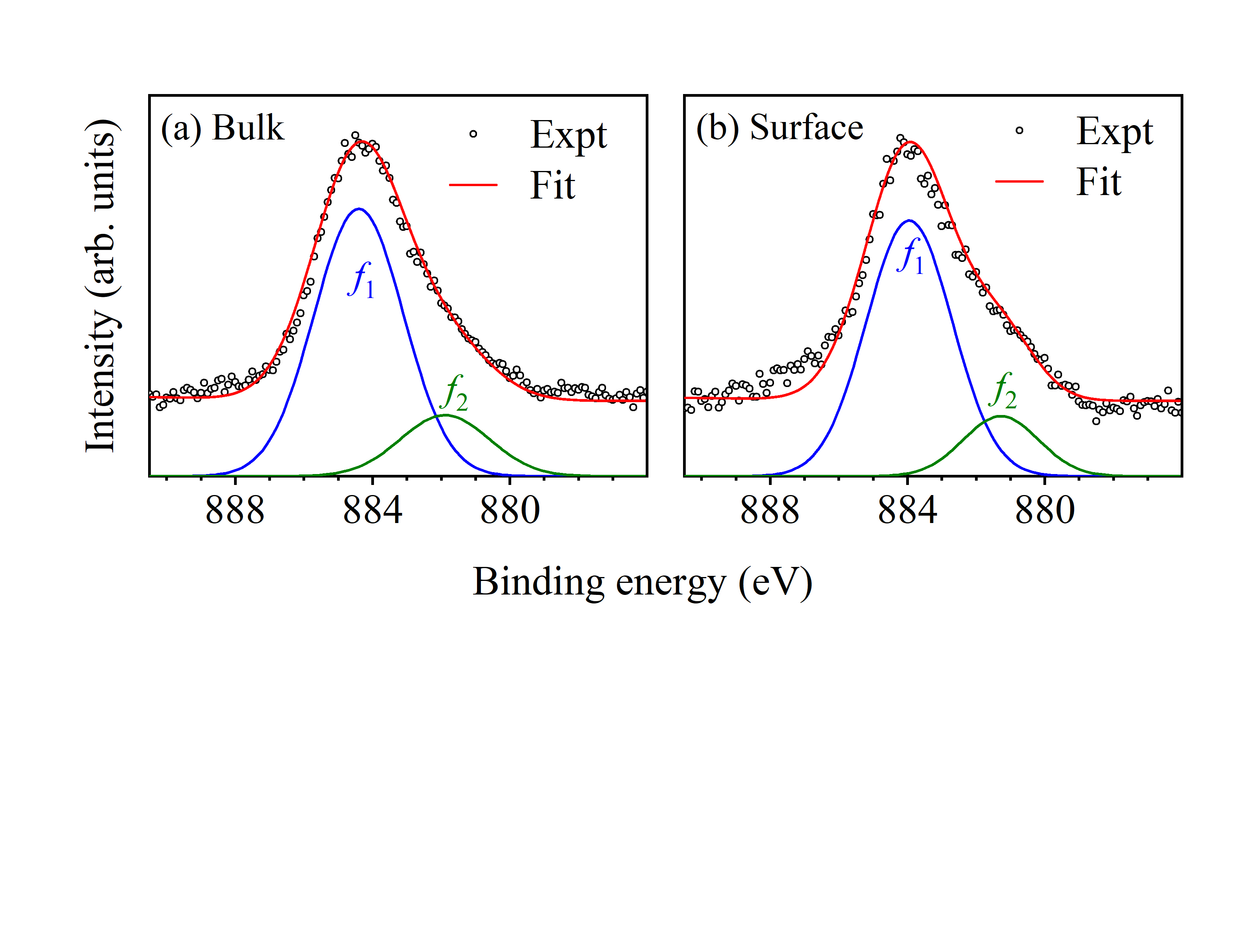}
\vspace{-20ex}
\caption{Simulated Ce 3$d_{5/2}$ spectra of CeCuAs$_2$ using Imer and Wuilloud's approach of GS calculation for (a) the bulk and (b) the surface (b) spectrum. Line superimposed over the experimental data is the simulated curve. The component peaks for the $f_1$ and $f_2$ final states are shown in the bottom panel.}
\label{Fig4-IWcalcn}
\end{figure}
	
In order to verify the above assertions with the existing models for such systems, we have simulated the Ce 3$d_{5/2}$ spectrum using Imer and Wuilloud's approach \cite{IW}, which is a simplified version of the Gunnarsson and Sch\"{o}nhammer model and considers only the configurations without their multiplet splittings. The calculated plots are shown in Fig. \ref{Fig4-IWcalcn}(a) (bulk) and (b) (surface) by the red line superimposing over the experimental data (black open circle). Here, the experimental spectra are obtained by subtracting the contributions due to b$_1$ and s$_1$ signals. The $f_2$ and $f_1$ final states are shown by the green and blue lines, respectively. The simulated results show an excellent representation of the experimental spectra for the following set of parameters. For bulk: on-site Coulomb repulsion strength, $U_{ff}$ = 8 eV, the core-hole potential $U_{fc}$ = 8.5 eV, energy of the unhybridized Ce 4$f$ level $e_f$ = -1.5 eV and the hybridization between the final states $f_0$, $f_1$ and f$_2$ states, denoted by $\Delta$ which is equal  to 0.55 eV. Using this calculation we got the value for the $f$ occupancy of 0.92 for the bulk spectrum. The simulation of the surface spectrum requires slightly different $e_f$ (= -1.7 eV) keeping all the other parameters same. The $f$-occupancy for surface Ce is found to be 0.94.
	
We note here that the Ce 3$d$ spectra analysed in Fig. \ref{Fig4-IWcalcn} is similar to the reported Ce 3$d$ spectral lineshape for polycrystalline CeCuAs$_2$\cite{chainani}. It appears clear that the features, b$_1$ and s$_1$ have a different origin and they are revealed in the experimental data from single crystalline samples. Based on the discussion above and published literature, we assign these features to the well screened final states where the holes created at the ligand sites and/or conduction band are further stabilized via formation of singlets with a Cu 3$d$ hole and/or delocalization as observed in other materials \cite{sawani,sawani2}. Similar effect is reported in Cu, Fe and Mn-based systems too \cite{A1_Zhang, A1_bos, A1_manganite, A1_ram}.

\begin{figure}
\centering
\includegraphics[width=0.5\textwidth]{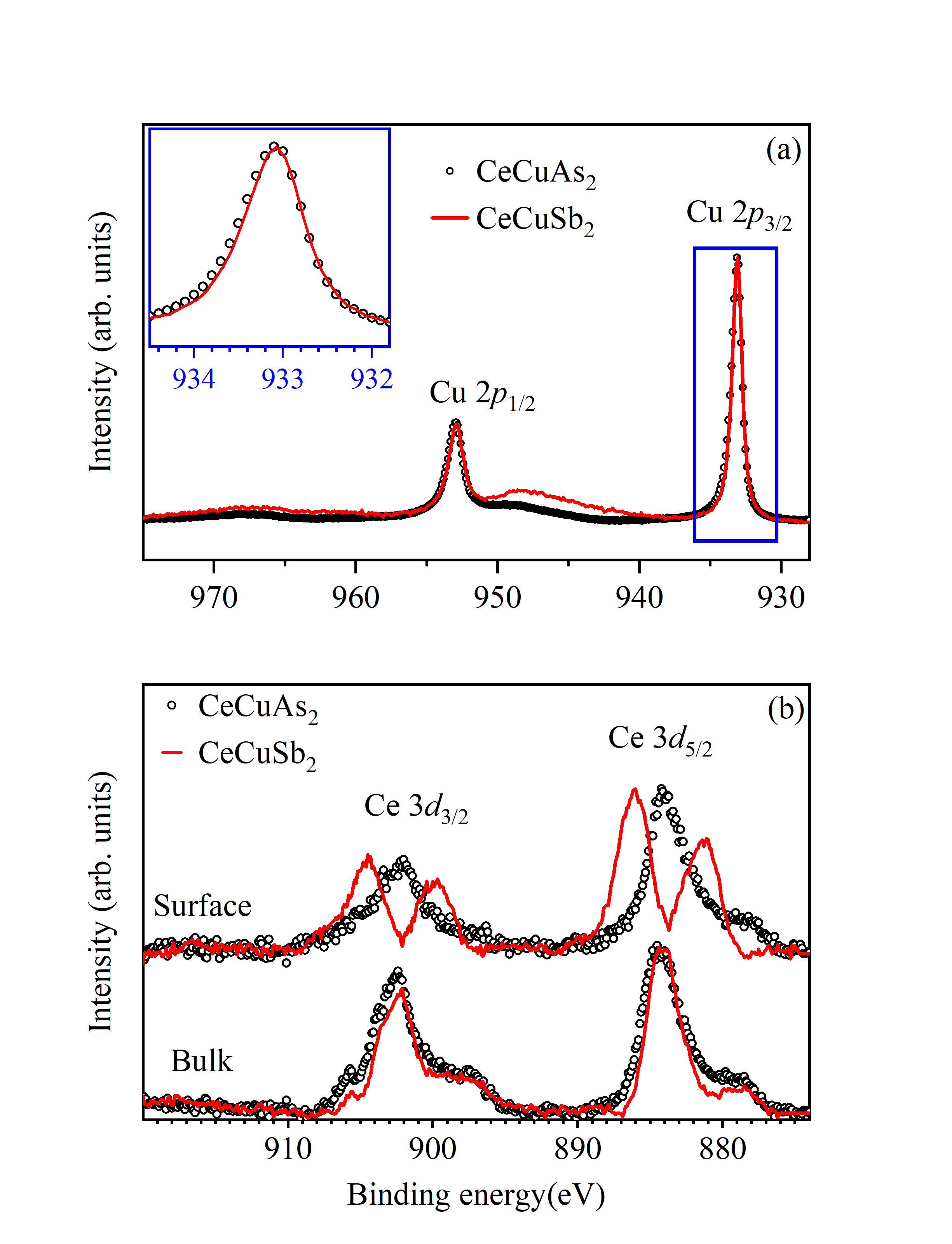}
\vspace{-2ex}
\caption{(a) Cu 2$p$ spectra of CeCuAs$_2$ ((open circles) and CeCuSb$_2$ (solid line) collected using 6 keV photon energy. Inset shows the boxed part of the spectra. (b) Comparison of the Ce 3$d$ surface, bulk spectra of CeCuAs$_2$(open circles) and CeCuSb$_2$(solid line).}
\label{Fig5-comp}
\end{figure}

In order to study the underlying scenario further, we compare the results of CeCuAs$_2$ with those of CeCuSb$_2$ in Fig. \ref{Fig5-comp}. Cu 2$p$ spectra of CeCuAs$_2$ (open circles) and CeCuSb$_2$ (solid line) are shown in Fig. \ref{Fig5-comp}(a); in the case of CeCuSb$_2$, Sb 3$s$ contributions are subtracted for one to one comparison. The lineshape of the main peak appear similar in both the cases except the fact that the spectral asymmetry is slightly larger in CeCuAs$_2$. This is shown with clarity in the inset of Fig. \ref{Fig5-comp}(a). The plasmon features are more intense in CeCuSb$_2$ presumably due to the highly conductive nature of CeCuSb$_2$ among these two materials.
	
The surface and bulk Ce 3$d$ spectra are compared in Fig. \ref{Fig5-comp}(b). The surface spectrum of CeCuSb$_2$ look very different from that of CeCuAs$_2$. The electronic structure of Sb1 at the terminated surface are significantly different from the other Sb-layers \cite{sawani}. Such difference for the pnictogen layers is less evident in CeCuAs$_2$. This might be a reason for such a different Ce 3$d$ surface spectra. The bulk Ce 3$d$ spectra, however, appear similar in both the cases. The lowest binding energy feature, s$_1$ is almost absent in CeCuSb$_2$ surface spectra while that in CeCuAs$_2$ is quite significant. In the bulk spectra too the intensity of the feature, b$_1$ in CeCuAs$_2$ is stronger than that in CeCuSb$_2$.

To investigate this further, we compare various bond lengths in these two compounds which is directly linked to the hybridization parameters. The Ce-Cu bondlength is 3.2992 \AA\ in CeCuAs$_2$ and 3.3615 \AA\ in CeCuSb$_2$. Shorter bondlength in CeCuAs$_2$ suggests stronger Ce-Cu hybridization compared to that in CeCuSb$_2$. On the other hand, the Cu-X2 bond length is smaller in CeCuSb$_2$ (2.6686 \AA) than in CeCuAs$_2$ (2.7718 \AA). Hence, the X2-Cu-X2 layer is more tightly packed in CeCuSb$_2$. Thus, the holes created in the conduction band due to Ce 3$d$ core-hole screening will have higher probability to propagate to the Cu-site in CeCuAs$_2$ than in CeCuSb$_2$ and form a singlet with the Cu 3$d$-hole. While these considerations provide a qualitative picture of the scenario, we hope future theoretical studies would help to establish the properties of this feature better.
	
\begin{figure}
\centering
\includegraphics[width=0.5\textwidth]{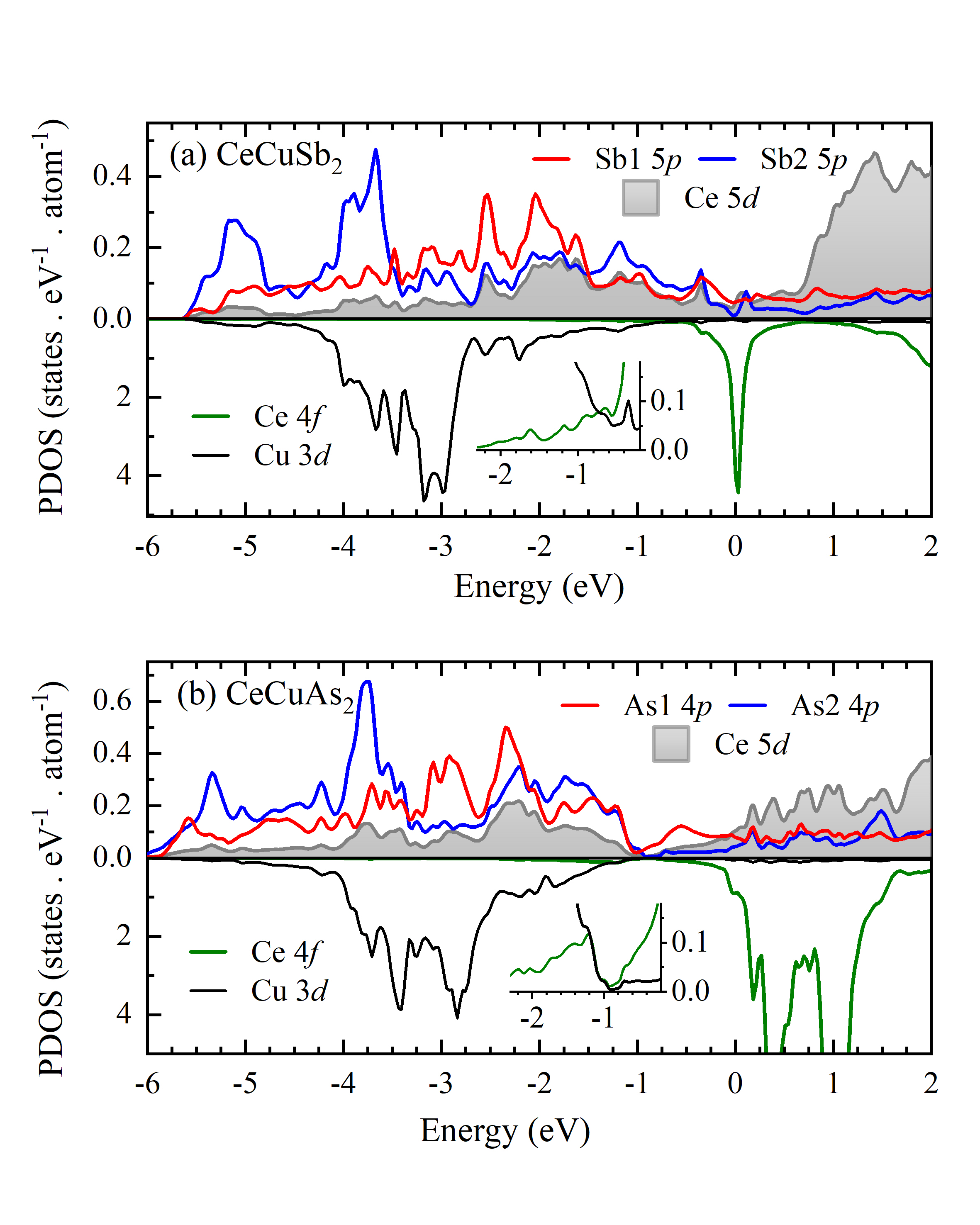}
\caption{Calculated partial density of states (PDOS) for (a) CeCuSb$_2$ and (b) CeCuAS$_2$. Insets show the Ce 4$f$ and Cu 3$d$ PDOS in an expanded PDOS scale.}
\label{Fig6-DFT}
\end{figure}

In Fig.~\ref{Fig6-DFT}, we have shown the calculated density of states (DOS) of CeCuSb$_2$ (Fig.~\ref{Fig6-DFT}(a)) and CeCuAs$_2$ (Fig.~\ref{Fig6-DFT}(b)). In CeCuSb$_2$, the major contribution in the bonding energy bands (-2.6 eV to -5.6 eV) are coming from the Cu 3$d$  and Sb2 5$p$ partial density of states (PDOS). Sb1 5$p$ contributions appear at higher energies compared to Sb2 5$p$ PDOS. The energy region between -3 eV and the Fermi level are contributed by Sb 5$p$ (both Sb layers), Ce 5$d$ and Cu 3$d$ states. Evidently, the hybridization of Sb2 5$p$ states with Ce 5$d$ and Cu 3$d$ PDOS is significant. The Ce 4$f$ PDOS primarily appear in the vicinity of the Fermi level, $\epsilon_F$. Contribution of Ce 4$f$ is quite low in the lower energy region, where Cu 3$d$ PDOS dominate; this is shown with clarity in the inset of Fig. \ref{Fig6-DFT}(a). Large intensity in the vicinity of $\epsilon_F$ suggests highly metallic character of the material.

On the other hand, Cu 3$d$ contributions appear much closer to the Fermi level in CeCuAs$_2$. PDOS at about -1 eV appears to be negligible indicating signature of an energy gap there. Ce 5$d$ contributions are shifted towards lower energy leading to a larger hybridization with the Cu 3$d$ states. Ce 4$f$ states also contribute significantly between -1 to -2 eV suggesting relatively larger Ce - Cu hybridization as also predicted from the bondlength analysis. The difference between As1 and As2 5$p$ states are significantly reduced in CeCuAs$_2$ compared to CeCuSb$_2$. These results supports the observations in the core level spectra such as slightly larger asymmetry of Cu 2$p$ feature in CeCuAs$_2$ arising from larger Cu 3$d$ PDOS near the Fermi level allowing larger low energy excitations. This also provides signature of larger Ce 4$f$ - Cu 3$d$ hybridization which might be a reason for larger intensity of b$_1$ in CeCuAs$_2$.
	
\begin{figure}
\centering
\includegraphics[width=0.5\textwidth]{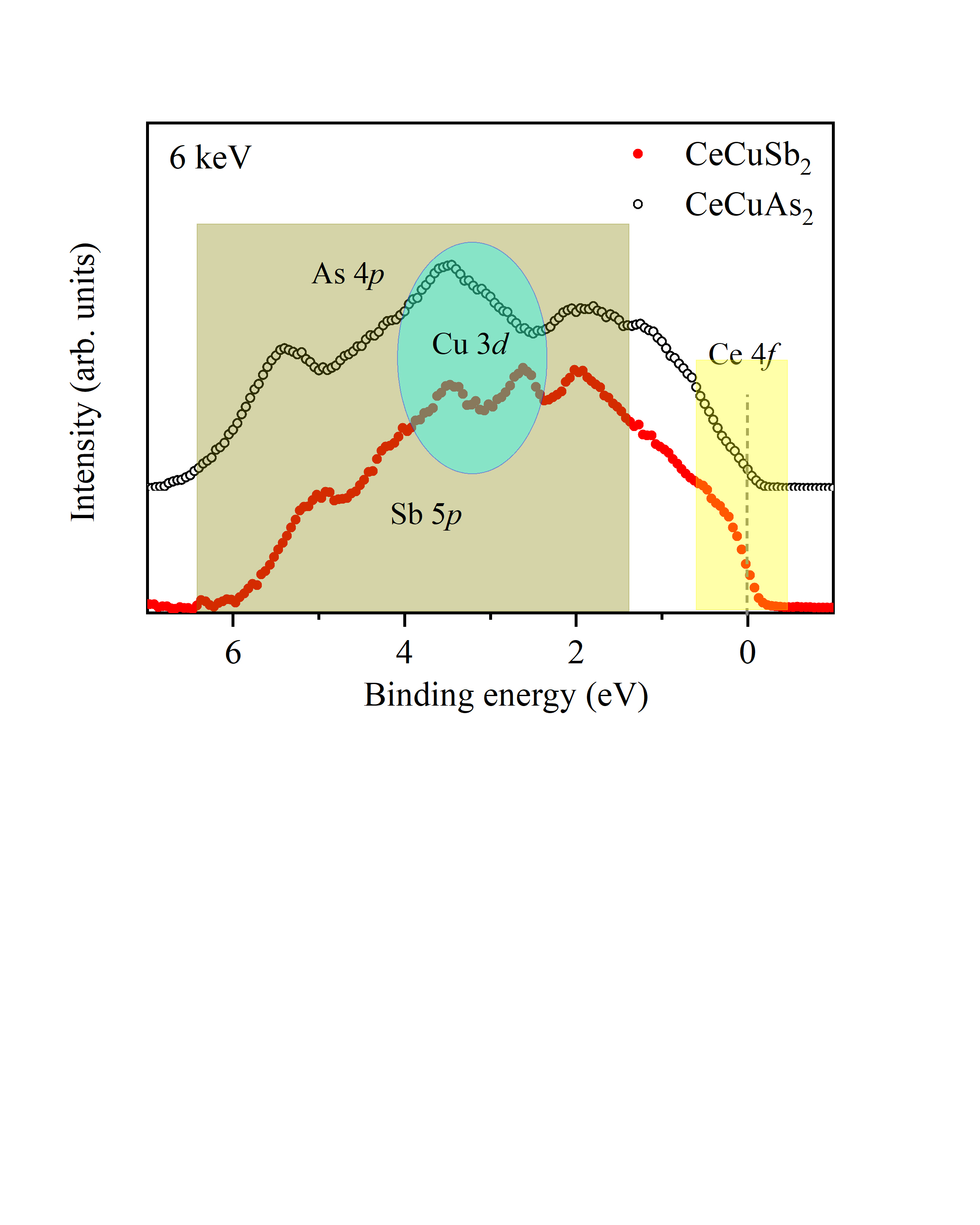}
\vspace{-32ex}
\caption{Valence band spectra of CeCuAs$_2$ and CeCuSb$_2$ at 6 keV and 45 K temperature. Different spectral regions based on band structure results are shown by shaded areas. The data show negligible intensity at the Fermi level in CeCuAs$_2$ while it is quite intense in CeCuSb$_2$.}
\label{Fig7-VB}
\end{figure}

In Fig.~\ref{Fig7-VB} we show the valence band spectra of CeCuSb$_2$ and CeCuAs$_2$ measured using 6 keV photon energy at 45 k sample temperature. Comparison of these data with the calculated results shown in Fig. \ref{Fig6-DFT} suggests that the binding energy region between 1.5 eV and 6 eV is primarily contributed by As 4$p$ / Sb 5$p$ states. Cu 3$d$ contributions appear in the vicinity of 3 eV binding energy. The peak at about 5.5 eV is contributed by As2 4$p$ states while the Sb2 5$p$ contributions appear at slightly lower binding energy as also observed in the calculated data. Cu 3$d$ contributions appear to be much broader in CeCuAs$_2$ compared to CeCuSb$_2$ consistent with their PDOS. The energy region close to $\epsilon_F$ is primarily contributed by Ce 5$d$ and 4$f$ PDOS with which are strongly hybridized with the As/Sb $p$ states. Interestingly, the spectral intensity at $\epsilon_F$ is quite significant in CeCuSb$_2$ suggesting highly metallic character of this material while the intensity in CeCuAs$_2$ is negligibly small. This is consistent with the observation of pseudogap in earlier studies \cite{chainani}.

\begin{figure}
\centering
\includegraphics[width=0.5\textwidth]{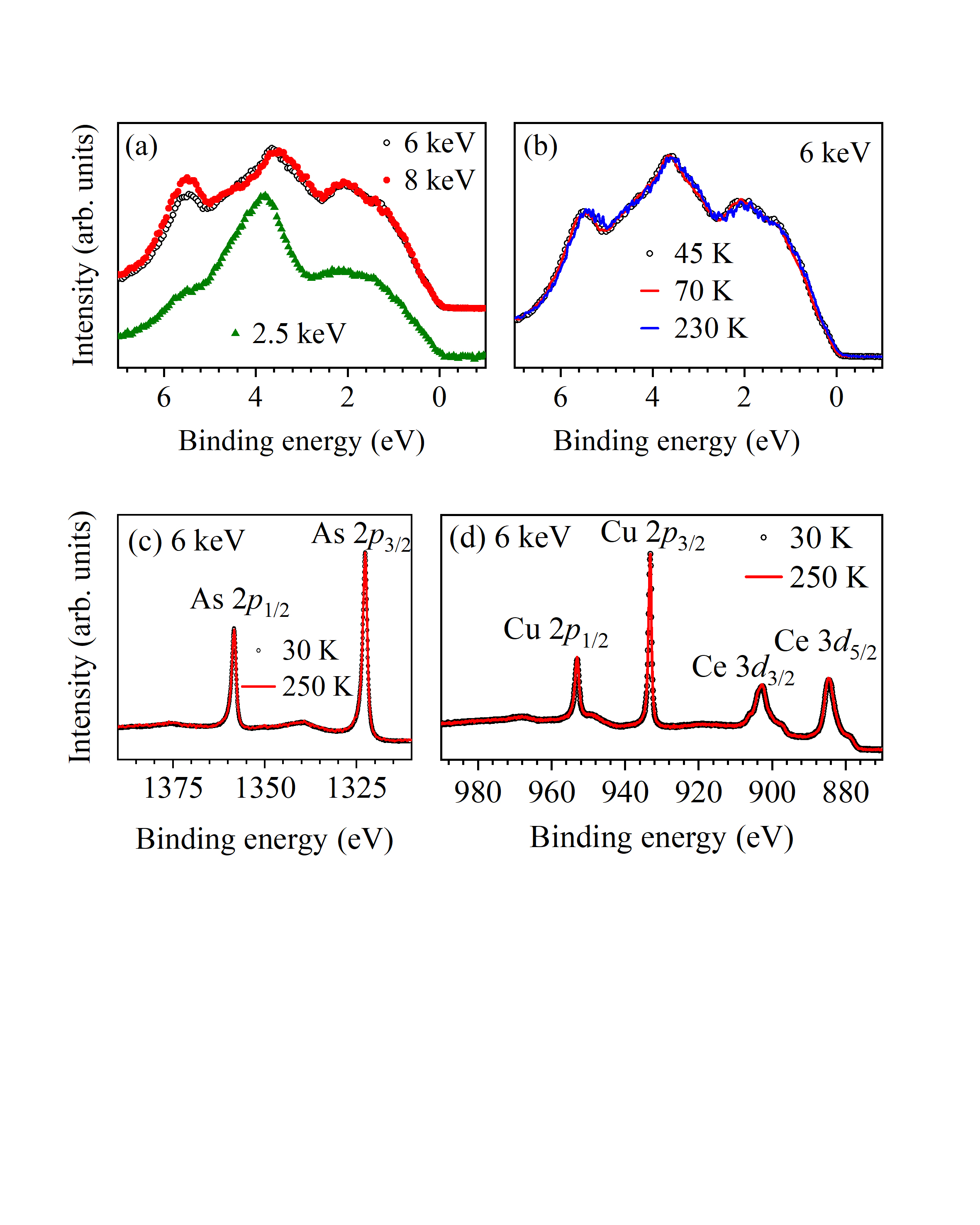}
\vspace{-24ex}
\caption{(a) Valence band spectra of CeCuAs$_2$ at 45 K using 2.5 keV (triangles), 6 keV (open circles) and 8 keV (solid circles) photon energies. (b) 6 keV valence band spectra at temperatures 45 K, 70 K and 230 K. (c) As 2$p$ spectra at 30 K and 250 K using 6 keV photon energy. (d) Cu 2$p$ and Ce 3$d$ spectra at 30 K and 250 K using 6 keV photon energy. The data at different temperatures show almost identical lineshape in all the cases.}
\label{Fig8-Temp}
\end{figure}
	
In Fig. \ref{Fig8-Temp}(a), we show the experimental valence band spectra collected at 45 K at different photon energies. The spectra collected using 6 keV and 8 keV photon energies overlap well except the intensity of the feature around 5.5 eV which is stronger in the 8 keV data. In the 2.5 keV data, the intensity of the feature around 4 eV enhances drastically compared to the intensity of the other features. The feature at 5.5 eV becomes much smaller in this spectrum. From the atomic cross-section data \cite{Yeh}, we observe that the cross section of Cu 3$d$ at 1486.6 eV photon energy is about an order of magnitude higher than the cross section of As 4$p$, Ce 4$f$ and 5$d$ states. The cross section reduces by 2 orders of magnitude at 8 keV photon energy in every case except Ce 4$f$ which reduces by 3 orders of magnitude. Thus, the change in intensity between 6 keV and 8 keV can be attributed essentially to the change in surface sensitivity and the feature at 5.5 eV constituted by As2 4$p$ PDOS is a bulk feature. The feature at 4 eV is primarily contributed by the Cu 3$d$ states. The intensity at $\epsilon_F$ is weak in every case indicating a semi-metallic behavior of this material.

We now investigate the evolution of the spectral functions with temperature. The spectra collected using 6 keV photon energy at different temperatures are shown in Fig. \ref{Fig8-Temp}(b), (c) and (d) for valence band, As 2$p$, Cu 2$p$ and Ce 3$d$ photoemissions. We do not observe any distinct change in the spectral lineshape with temperature in any of the cases. This suggests that the thermal effect on the electronic structure of this system in the temperature range studied is not significant enough to be detected by these experiments.

\section{Conclusion}
	
In conclusion, we have investigated the electronic structure of ternary CeCu-pnictides (CeCuX$_2$; X= Sb and As) employing depth-resolved hard $x$-ray photoemission spectroscopy. We observe that these materials cleaved at the pnictogen layers leaving the terminated surface dominated by the pnictogen atoms forming a square-net structure. This has significant implication in the study and application of these materials as the square-net structure is known to host topologically protected states. In the present scenario, such states can be produced at the surface as well as at the X1 layer within the bulk of the material. CeCuAs$_2$ is an unique Kondo material which does not show magnetic order down the lowest temperature studied while other materials in this class show magnetic ground state. Ce 3$d$ core level spectra do not show the signature of $f_0$ feature in both the surface and bulk cases. We observe several features in the Ce 3$d$ spectra suggesting strong hybridization between the Ce 4$f$/5$d$ states with the valence states. The feature at lowest binding energy of the Ce 3$d$ spectra is found to be strong; the intensity is relatively larger in CeCuAs$_2$ which is an indication of stronger Cu-Ce hybridization in CeCuAs$_2$. The valence band spectra is found to be consistent with the calculated results using density functional theory. The intensity close to the Fermi level show highly metallic ground state of CeCuSb$_2$ while CeCuAs$_2$ is semi-metallic. The spectral functions collected at different temperature show identical lineshape indicating that the temperature induced changes are subtle in these materials. Hybridization between various valence states are found to be complex and leads to significant change in the spectral functions, which might be responsible for the exoticity of this novel Kondo system.

\section{Acknowledgements}
Authors acknowledge the financial support under India-DESY program and Department of Atomic Energy (DAE), Govt. of India (Project Identification no. RTI4003, DAE OM no. 1303/2/2019/R\&D-II/DAE/2079 dated 11.02.2020). K. M. acknowledges financial support from BRNS, DAE, Govt. of India under the DAE-SRC-OI Award (grant no. 21/08/2015-BRNS/10977).


\begin{thebibliography}{11}

\bibitem{review} G. R. Stewart, Rev. Mod. Phys. {\bf 56}, 755 (1984); N. B. Brandt and V. V.
Moshchalkov, Adv. Phys. {\bf 33}, 373 (1984); F. Steglich, J. Magn. Magn. Mater {\bf 100}, 186 (1991).

\bibitem{Parlebas}
J. C. Parlebas, Phys. Stat. Sol. (b) \textbf{160} 11 (1990).

\bibitem{book}
A. C. Hewson, \emph{The Kondo Problem to Heavy Fermions} (Cambridge University Press, Cambridge, 1993).
%
\bibitem{patil-213bulk}
S. Patil, K. K. Iyer, K. Maiti and E. V. Sampathkumaran, Phys. Rev. B \textbf{77}, 094443 (2008).
%
S. Patil, S. K. Pandey, V. R. R. Medicherla, R. S. Singh, R. Bindu, E. V. Sampathkumaran, and K. Maiti, J. Phys.: Condens. Matter \textbf{22}, 255602 (2010);
%
S. Patil, G. Adhikary, G. Balakrishnan, and K. Maiti, J. Phys.: Condensed Matter \textbf{23}, 495601 (2011).
%
\bibitem{Sengupta1}
K. Sengupta, E. V. Sampathkumaran, T. Nakano, M. Hedo, M. Abliz, N. Fujiwara, Y. Uwatoko, S. Rayaprol, K. Shigetoh, T. Takabatake, Th. Doert, and J. P. F. Jemetio, Phys. Rev. B \textbf{70}, 064406 (2004).

\bibitem{CeCuX2-1}
H. Sprenger, J. Less-Common Met. \textbf{34}, 39 (1974);
%
J. Stepien-Damm, D. Kaczorowski, and R. Troc, J. Less-Common Met. \textbf{132}, 15 (1987);
%
M. Brylak, M. H. Moller, and W. Jeitschko, J. Solid State Chem. \textbf{115}, 305 (1995).
%
%
\bibitem{CeCuX2-2}
D. Rutzinger, C. Bartsch, M. Doerr, H. Rosner, V. Neu, Th. Doert, and M. Ruck, J. Solid State Chem. \textbf{183}, 510 (2010);
%
M. Szlawska and D. Kaczorowski, J. Alloys Compd. \textbf{451}, 464 (2008);
%
E. V. Sampathkumaran, K. Sengupta, S. Rayaprol, K. K. Iyer, Th. Doert and J. P. F. Jemetio, Phys. Rev. Lett. \textbf{91}, 036603 (2003).
%
\bibitem{thamizh}
S. Araki, N. Metoki, A. Galatanu, E. Yamamoto, A. Thamizhavel, and Y. Onuki, Phys. Rev. B \textbf{68}, 024408 (2003);
%
A. Thamizhavel, T. Takeuchi, T. Okubo, M. Yamada, R. Asai, S. Kirita, A. Galatanu, E. Yamamoto, T. Ebihara, Y. Inada, R. Settai, and Y. Onuki, \emph{ibid}. \textbf{68}, 054427 (2003).
%
\bibitem{patil_CEF}
S. Patil, V. R. R. Medicherla, R. S. Singh, S. K. Pandey, E. V. Sampathkumaran, and K. Maiti, Phys. Rev. B \textbf{82}, 104428 (2010);
%
K. Maiti, S. Patil, G. Adhikary, and G. Balakrishnan, J. Phys.: Conf. Ser. \textbf{273}, 012042(2011);
%
S. Patil, A. Generalov, M. Güttler, P. Kushwaha, A. Chikina, K. Kummer, T. C. R\"{o}del, A. F. Santander-Syro, N. Caroca-Canales, C. Geibel, S. Danzenb\"{a}cher, Yu. Kucherenko, C. Laubschat, J. W. Allen, and D. V. Vyalikh ,  Nat. Commun. \textbf{7}, 11029 (2016).
%
\bibitem{Weschke}
E. Weschke, C. Laubschat, R. Ecker, A. H\"{o}hr, M. Domke, G. Kaindl, L. Severin, and B. Johansson,
Phys. Rev. Lett. \textbf{69}, 1792 (1992).
%
\bibitem{Allen}
J. W. Allen, S. J. Oh, O. Gunnarsson, K. Sch\"{o}nhammer, M. B. Maple, M. S. Torikachvili, and I. Lindau, Adv. Phys. \textbf{35}, 275 (1986).
%
\bibitem{patil-SOC}
S. Patil,  V. R. R. Medicherla, R. S. Singh, E. V. Sampathkumaran, and K. Maiti, EPL \textbf{97}, 17004 (2012).
%
\bibitem{Gordon}
R. A. Gordon, M. G. Alexander, C. J. Warren, F. J. DiSalvo and R. P\"{o}ttgen, J. Alloys Compounds \textbf{248}, 24 (1997).

\bibitem{Chevalier}
B. Chevalier, P. Lejay, J. Etourneau, and P. Hagenmuller, SolidState Commun. \textbf{49}, 753 (1984).

\bibitem{Szlawska}
M. Szlawska, D. Kaczorowski, A. \'{S}lebarski, L. Gulay, and J. Stepie\'{n}-Damm, Phys. Rev. B \textbf{79}, 134435 (2009).

\bibitem{Sengupta2}
K. Sengupta S.Rayaprol, E.V.Sampathkumaran, Th. Doert, J.P.F.Jemetio, Physica B \textbf{348}, 465-474 (2004); E. V. Sampathkumaran, T. Ekino, R. A. Ribeiro, Kausik Sengupta, T. Nakano, M. Hedo, N. Fujiwara, M. Abliz, Y. Uwatoko, S. Rayaprol, Th. Doert, and J. P. F. Jemetio, Physica B \textbf{359-361} 108 (2005)

\bibitem{Blaha}
P. Blaha, K. Schwarz, G. K. H. Madsen, D. Kvasnicka, and J. Luitz, {\it WIEN2K: An Augmented lane Wave plus local Orbitals Program for Calculating Crystal Properties}. (Karlheinz Schwarz, Technische Universität Wien, Austria, 2001) ISBN 3-9501031-1-2 (2001)
		
\bibitem{Perdew}
John P. Perdew, Kieron Burke, and Matthias Ernzerhof, Phys. Rev. Lett. \textbf{77}, 3865 (1996).
		
\bibitem{Hohen}
P. Hohenberg and W. Kohn, Phys. Rev. \textbf{136}, 864-871 (1964).
		
\bibitem{escapedepth}
M. P. Seah and W.A. Dench, Surface and Interface Analysis, \textbf{1}, 2 (1979).
		
\bibitem{Yeh}
Yeh and Lindau, Atomic Data And Nuclear Data Tables \textbf{32}, l-l55 (1985).

\bibitem{sawani}
S. Datta, R. P. Pandeya, A. B. Dey, A. Gloskovskii, C. Schlueter, T. Peixoto, A. Singh, A. Thamizhavel, and K. Maiti, Phys. Rev. B \textbf{105}, 205128 (2022).
 		
\bibitem{chainani}
A. Chainani, M. Matsunami, M. Taguchi, R. Eguchi, Y. Takata, M. Oura, S. Shin, K. Sengupta, E. V. Sampathkumaran, Th. Doert, Y. Senba, H. Ohashi, K. Tamasaku, Y. Kohmura, M. Yabashi, and T. Ishikawa, Phys. Rev. B \textbf{89}, 235117 (2014).
		
\bibitem{Biswas}
D. Biswas, S. Thakur, G. Balakrishnan and K. Maiti, Sci. Rep. \textbf{5}, 17351 (2015).

\bibitem{surface-bulk}
K. Maiti, Priya Mahadevan, and D. D. Sarma, PRL \textbf{80}, 2885 (1998).
		
\bibitem{gun}
O. Gunnarsson and K. Schonhammer, Phys. Rev. B \textbf{28}, 4315 (1983).
		
\bibitem{nickelates_kbm}
K. Maiti, P. Mahadevan, and D. D. Sarma, Phys. Rev. B {\bf 59}, 12457 (1999).
		
\bibitem{RMP_Fujimori}
M. Imada, A. Fujimori, and Y.Tokura, Rev. Mod. Phys. \textbf{70}, 1039 (1998).
		
\bibitem{gun_f0}
O. Gunnarsson and K. Schonhammer, Phys. Rev. B \textbf{31}, 4815 (1985).
		
\bibitem{Laubschat}
C. Laubschat, E. Weschke, C. Holtz, M. Domke, O. Strebel, and G. Kaindl, Phys. Rev. Lett. \textbf{65}, 1639 (1990)

\bibitem{IW}
J. M. Imer and E. Wuilloud, Z. Phys. B.-Condensed Matter \textbf{66}, 153-160 (1987).

\bibitem{sawani2}
S. Datta, R. P. Pandeya, A. B. Dey, A. Gloskovskii, C. Schlueter, T. R. F. Peixoto, A. Singh, A. Thamizhavel, and K. Maiti, J. Phys.: Condens. Matter \textbf{35}, 235601 (2023).
		
\bibitem{A1_Zhang}
F. C. Zhang and T. M. Rice, Phys. Rev. B {\bf 37}, 3759(R) (1988).
		
\bibitem{A1_bos}
T. B\"{o}ske, K. Maiti, O. Knauff, K. Ruck, M. S. Golden, G. Krabbes, J. Fink, T. Osafune, N. Motoyama, H. Eisaki, and S. Uchida, Phys. Rev. B \textbf{57}, 138 (1998).
		
\bibitem{A1_manganite}
K. Horiba, M. Taguchi, A. Chainani \textit{et. al.}, Phys. Rev. Lett. \textbf{93}, 236401 (2004).
		
\bibitem{A1_ram}
R. P. Pandeya A. P. Sakhya, S. Datta, T. Saha, G. D. Ninno, R. Mondal, C. Schlueter, A. Gloskovskii, P. Moras, M. Jugovac, C. Carbone, A. Thamizhavel, and K. Maiti, Phys. Rev. B {\bf 104}, 094508 (2021).
		
\end{thebibliography}
\end{document}